\documentclass[aps,prl,reprint,superscriptaddress]{revtex4-1}
\usepackage{blindtext}
\usepackage{graphicx}
\usepackage{amsmath}
\usepackage{multirow}
\usepackage{url}
\usepackage{notes2bib}
\usepackage[colorlinks,
            linkcolor=blue,
            anchorcolor=blue,
            citecolor=blue,
            ]{hyperref}
\usepackage{setspace}

\begin{document}

\title{Wave scattering properties of multiple weakly-coupled complex systems}

\author{Shukai Ma}
\email{skma@umd.edu}
\affiliation{Department of Physics, University of Maryland, College Park, Maryland 20742-4111, USA}
\author{Bo Xiao}
\affiliation{Department of Electrical and Computer Engineering, University of Maryland, College Park, Maryland 20742-3285, USA}
\author{Zachary Drikas}
\affiliation{U.S. Naval Research Laboratory, Washington, DC 20375, USA}
\author{Bisrat Addissie}
\affiliation{U.S. Naval Research Laboratory, Washington, DC 20375, USA}
\author{Ronald Hong}
\affiliation{U.S. Naval Research Laboratory, Washington, DC 20375, USA}
\author{Thomas M. Antonsen}
\affiliation{Department of Physics, University of Maryland, College Park, Maryland 20742-4111, USA}
\affiliation{Department of Electrical and Computer Engineering, University of Maryland, College Park, Maryland 20742-3285, USA}
\author{Edward Ott}
\affiliation{Department of Physics, University of Maryland, College Park, Maryland 20742-4111, USA}
\affiliation{Department of Electrical and Computer Engineering, University of Maryland, College Park, Maryland 20742-3285, USA}
\author{Steven M. Anlage}
\affiliation{Department of Physics, University of Maryland, College Park, Maryland 20742-4111, USA}
\affiliation{Department of Electrical and Computer Engineering, University of Maryland, College Park, Maryland 20742-3285, USA}


\begin{abstract}
The statistics of scattering of waves inside single ray-chaotic enclosures have been successfully described by the Random Coupling Model (RCM). We expand the RCM to systems consisting of multiple complex ray-chaotic enclosures with variable coupling scenarios. The statistical properties of the model-generated quantities are tested against measured data of electrically large multi-cavity systems of various designs. The statistics of model-generated trans-impedance and induced voltages on a load impedance agree well with the experimental results. The RCM coupled chaotic enclosure model is general and can be applied to other physical systems including coupled quantum dots, disordered nanowires, and short-wavelength electromagnetic propagation through rooms in buildings, aircraft and ships.

\end{abstract}

\maketitle

\section {I. introduction}

It is of interest to study the scattering properties of complex ray-chaotic systems in the semi-classical limit. Examples include atomic nuclei \cite{Haq1982}, quantum dot transport \cite{Alhassid2001}, and the flow of electromagnetic (EM) waves through electrically-large resonant systems \cite{Doron1990,So1995,Holloway2003,Kuhl2005,Hoijer2013,Gifuni2016}. Concatenating two or more such systems is also of great interest, but not as extensively studied. 
Such coupled cavity systems can be realized in a wide range of physical platforms from inter-connected photonic crystal cavities \cite{Notomi2008}, to Cooper pair boxes in superconducting resonators \cite{Wallraff2004}, to microwave (MW) billiards \cite{Alt1998}, and others. It has proven possible to perform experiments for interacting systems and measure transmission as a function of coupling. Examples include measurements of conductance of quantum dots systems with coupled electron billiards \cite{Porod1992,Waugh1995,Brunner2008}, resistance of disordered nanowires modeled by a cascade of quantum dots \cite{Chalker1988,Sau2012, Beenakker2015, Zhang}, and simulating resonance strength of coupled quantum mechanical systems with superconducting MW billiards \cite{Dietz2006}, etc. 
Likewise, the EM wave properties of inter-connected electrically large enclosures, like the power flow and the impedance or scattering parameters, are also widely studied in engineering \cite{Tait2011,Tait2011a} in situations ranging from computer enclosures to rooms or buildings \cite{Junqua2005,Kaya2009, Hemmady2012,Gagliardi2015}. These settings are typically found to be ray-chaotic and highly over-moded, posing challenges to both numerical and experimental analysis means. A ray-chaotic enclosure has the property that two rays launched with slightly different initial conditions will separate exponentially with time as they continue bouncing inside the structure \cite{Ott2007,Dietz2016}. On the other hand, a minute change of the structure boundary condition can drastically affect the pre-established field profile inside the system \cite{Dupre2015, Kaina2015, DelHougne2018}. Though deterministic approaches are available \cite{Parmantier2004}, the chaotic properties make the numerical solutions vulnerable to small changes and uncertainties of interior structure details.

In the situations just described statistical and/or approximate approaches can be more useful than deterministic methods (e.g., direct numerical computations for a specific configuration). 
Examples include the Baum-Liu-Tesche electromagnetic topology approach in which the system is separated into sub-volumes and waves travelling between the sub-volumes are computed \cite{baum1978analysis}. In the Power Balance Method the mean power flow through connected over-moded cavities is calculated \cite{Junqua2005, Hill1994, Hill1998}. The Dynamic Energy Analysis (DEA) method involves solving for the phase-space energy density on a gridded domain \cite{Lyon1995}, and describes mean high-frequency wave energy distributions in all sub-systems \cite{Tanner2009, Bajars2017}. The Random Coupling Model (RCM) determines the statistical properties of the impedance and scattering parameters for complex enclosures \cite{Wigner1955,Zheng2006,Zheng2006a,Hart2009,Yeh,Gradoni2014,Xiao2016}. In contrast to the other above mentioned methods, the RCM generates both mean-field and statistical predictions, treats interference and utilizes a minimum of information, namely, the system coupling details and the enclosure loss parameter \cite{Zheng2006b,Gradoni2012,Li2015,Fan2017,Luyao2017}.
It was recently demonstrated that two single RCM enclosures with a specific scaling relationship with regard to size, frequency and wall conductivity share the same normalized impedance statistics \cite{Xiao2018}. This work paves the way for experimentally testing the wave properties of an extensive variety of networks of large coupled complex systems experimentally in a typical lab environment by studying their scaled-down-in-size counterparts \cite{Hill2009,GilGil2016}. 

It has been conjectured that the statistical properties of quantum spectra of classically chaotic systems can be described by Random Matrix Theory (RMT) \cite{McDonald1979,Casati1980,Berry1981,Bohigas1984,Heusler2007}. In particular, RMT can be applied to wave chaotic systems in the short wavelength regime.
The Heidelberg approach \cite{Verbaarschot1985} describes wave scattering from a highly over-moded complex system connected to the outside world by a finite number of scattering channels \cite{Fyodorov2004,Fyodorov2010,Mitchell2010,Fyodorov2012,Kumar2017}. RMT based studies for over-moded cavities with a single channel and multiple channels were treated in Ref. \cite{Zheng2006a,Warne2003,Hemmady2005}. The treatment of apertures as multiple correlated channels was formulated in Ref. \cite{Gradoni2015}.


We describe in this manuscript our efforts: (1) to experimentally investigate electrically large coupled enclosure systems utilizing both `full-scale' and `miniaturized' over-moded electromagnetic cavities, (2) to study the propagation of waves in such systems under a variety of conditions, by varying single cavity loss, inter-cavity coupling strength, and the total number of connected cavities, (3) to apply the Random Coupling Model to the multi-cavity system and compare the theoretical predictions with the measured data. The paper is organized as follows: in section II, we introduce the experimental set-up and discuss the measured results; in section III, we introduce a statistical model to study such cavity cascade systems based on an extension of the RCM; in section IV, we compare the predicted impedance and induced voltage statistics with measured data for both scaled and full scale cavity cascade set-ups. We summarize and discuss future directions in section V.

\section {II. Experimental set-up}
\begin{figure}
\includegraphics[width=0.45\textwidth]{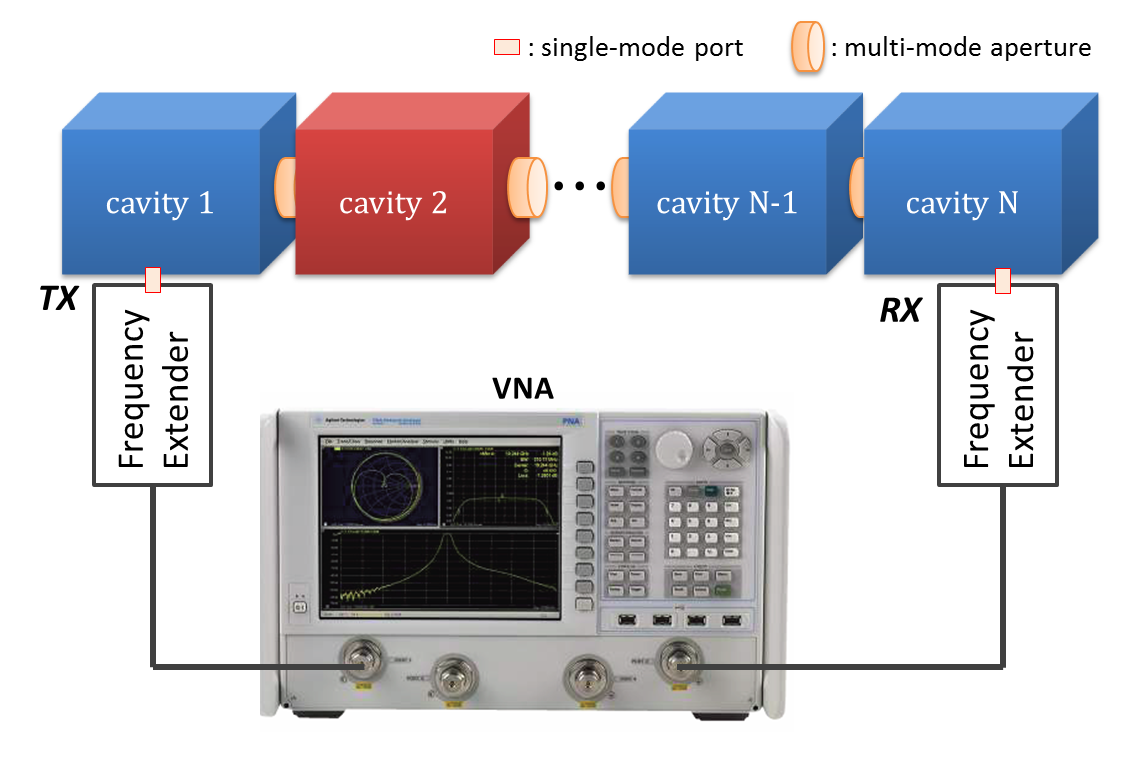}
\caption{\label{fig:VNA} Schematic of the cavity cascade experimental set-up. $N$ identical cavities are connected in a linear chain with aperture connections. Single-mode ports are installed at the first and last cavities in the chain. The scattering properties are measured with a VNA and frequency extenders. Independently controlled mode stirrers are located in each cavity.}
\end{figure}

Here we describe the basic topology of the coupled cavity system, a linear cavity chain connected through apertures as shown schematically in Fig. \ref{fig:VNA}. The cavities are electrically large and have an irregular interior structure on length scales greater than the wavelength of the EM waves that are sent through the system. For the full scale experiment, EM waves from 3.95-5.85GHz are injected into cavities of dimension $0.762\times0.762\times1.778 m^3$ through WR178 band single-mode waveguides (Fig. \ref{fig:connec}). Correspondingly, the $\times20$ scaled down version has the dimension of $0.038\times0.038\times0.089 m^3$ and is fed by EM waves from 75-110GHz through WR10 single-mode waveguides from a Virginia Diodes Vector Network Analyzer (VNA) Extenders (Fig. \ref{fig:VNA} and Fig. \ref{fig:connec}). The cavities contain mode stirrers of irregular shape to create many realizations of ray chaos in their interior.
With the scaled-down dimension and scaled-up operating frequency, identical statistical electrical properties are achieved in the two configurations \cite{Xiao2018}.

Apertures are created between the cavities to establish a controlled degree of inter-cavity coupling. In the scaled cavity case, rectangular and circular shaped apertures are employed between nearest neighbor cavities in the chain. 
The size and shape of the apertures are such that when the transverse fields in the apertures are represented in a basis of the modes of a waveguide (with equivalent cross section), either 5 or 100 of these modes (for the rectangular or circular apertures, respectively) would be above cut-off (propagating) at 110 GHz.
In the full scale cases, circular shaped apertures are adopted which allow 100 propagating modes at 5.85GHz.
The area of both sets of apertures are small compared with the surface area of the cavity to ensure a reasonably well-defined single cavity volume ($A_{aper}/A_{cav} \approx 0.78\%$ for a one-circular-aperture cavity, where $A_{cav}, A_{aper}$ are the cavity inner wall area and the aperture surface area).
The total number of cavities making up the linear chain is varied from one to three. In order to create a large ensemble for statistical analysis, independent mode stirrers are employed inside each individual cavity \cite{Frazier2013,Frazier2013a,Hemmady,Drikas2014}. Single-mode waveguide ports are created on the first and last cavity in the cascade.
The $2\times2$ Scattering(S)-parameters of the entire cavity cascade system are measured with the VNA. 
The initial positions of the mode stirrers are randomly assigned. We then rotate the stirrers by the same increment per step. 
The measurement cases described in this paper are summarized in Table. \ref{tab:catalog}.

The amount of loss in a cavity is controlled by varying the temperature of the copper walls in the scaled cavities \cite{Xiao2018}, and by placing RF absorber cones in the full scale set-ups. The single cavity `lossyness' is described by the RCM loss parameter $\alpha$, which is defined as the ratio of the 3-dB bandwidth of a typical resonance mode to the mean frequency spacing between the modes \cite{Gradoni2014, Xiao2018}. The loss parameter can be explicitly expressed as $\alpha = k^2/(Q\,\Delta k_n^2)$, where $k$ is the wavevector, and $Q$ and $\Delta k_n^2$ are the quality factor and averaged mode spacing for modes near $k$.
At room temperature the loss parameter of a scaled and a full scale cavity can both be made equal at a value of $\alpha \sim 9$. By matching the single cavity loss and the scaling in Maxwell equations, equivalent EM wave statistical properties are expected between the two experimental set-ups \cite{Xiao2018}.  

\begin{table*}
\begin{ruledtabular}
\begin{tabular*}{\textwidth}{@{\extracolsep{\fill}}llcc}
$N_{cav}$ & $\alpha$ & Dimension & Aperture (Coupling strength)\\
\hline
1 & 9.1 & scaled & N/A\\
2 and 3 & 9.1 & scaled & rectangular ($10^{-4}$)\\
2 and 3 & 9.1 & scaled & circular ($10^{-2}$)\\
1 & 5.7, 7.5 and 9.7 & full scale & N/A\\
2 and 3 & 5.7, 7.5 and 9.7 & full scale & circular ($10^{-2}$)\\
\end{tabular*}
\end{ruledtabular}
\caption{\label{tab:catalog} Measurement cases employed in this manuscript. $N_{cav}$ and $\alpha$ refer to the total number of cavities in the cascade chain and the single cavity loss parameter. The physical scale and the type of the aperture between cavities are described in the columns labeled `Dimension' and `Aperture'.}
\end{table*}

\begin{figure}
\includegraphics[width=0.5\textwidth]{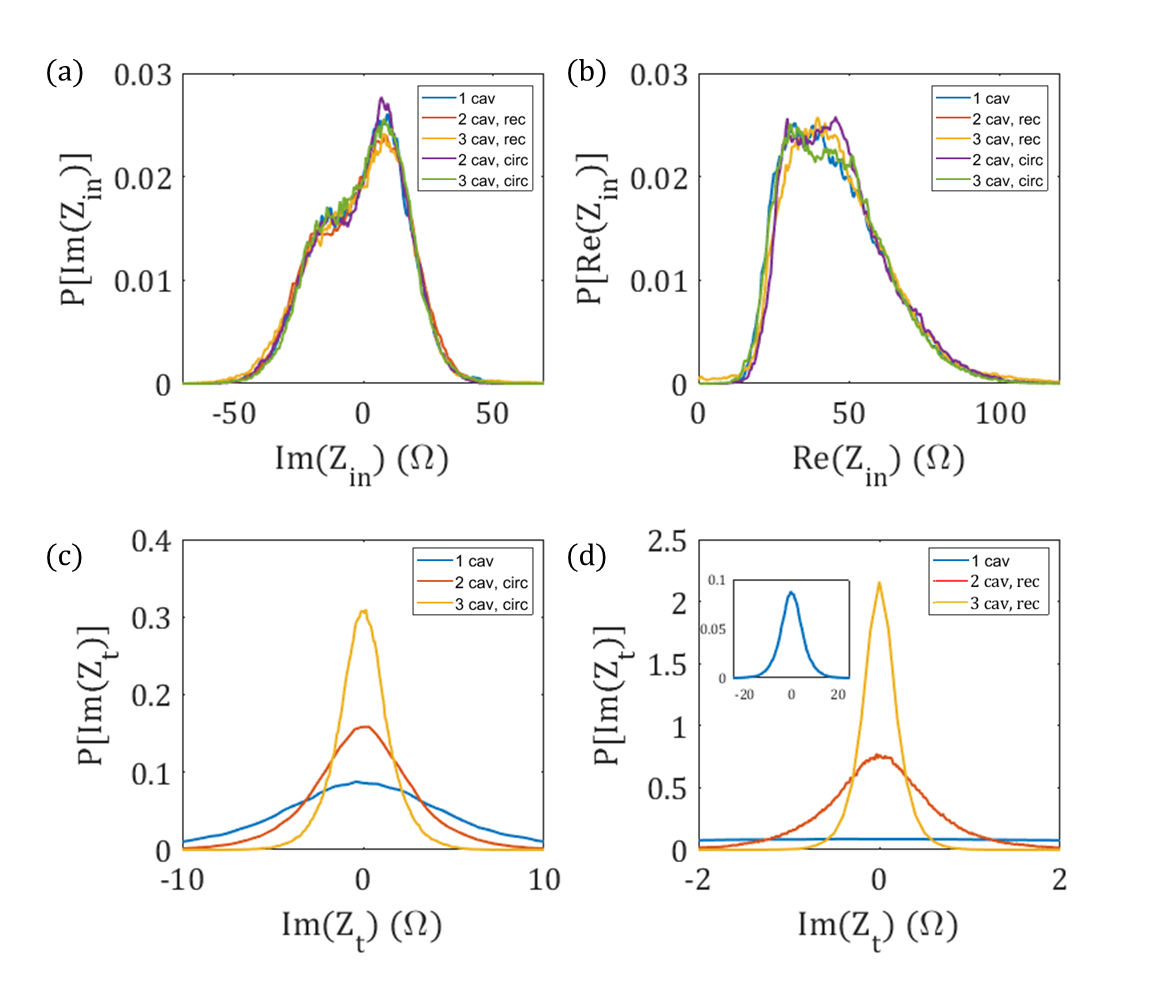}
\caption{\label{fig:Zinandt} Statistical analysis of the measured impedances $Z_{in}$ and $Z_{t}$ of the scaled cavity cascade systems. (a) and (b): the PDFs of imaginary and real part of the measured $Z_{in}$. (c) and (d): the PDFs of $Z_t$ imaginary parts from 1-3 cavity cascade system with circular and rectangular aperture connections, respectively.
The inset in (d) is the complete 1 cavity $Im(Z_t)$ PDF curve.}
\end{figure}

We first present the measured impedance statistics of the scaled cavity system with different connected cavity number and aperture coupling as shown in Fig. \ref{fig:Zinandt}. The small-in-size character of the scaled system allows us to easily create various cavity connection scenarios and change the inter-cavity coupling. The quantities of interest as plotted in Fig. \ref{fig:Zinandt} are the PDFs of the raw input impedance $Z_{in}$ and trans-impedance $Z_{t}$, defined as:
\begin{equation}
    Z_{in}=\frac{U_{TX}}{I_{TX}}, \quad\quad Z_{t}=\frac{U_{RX}}{I_{TX}}
\end{equation}
where TX(RX) refers to the transmitting(receiving) port, and U (I) is the complex phasor voltage (current) at the single-mode ports. As shown in Fig. \ref{fig:Zinandt} (a) and (b), the statistics of the measured $Z_{in}$ remain essentially unchanged as the total number of connected cavities or the aperture coupling are varied. 
The cavity coupling strength is defined as the ratio between the total number of propagating aperture modes and the total number of cavity modes at a given frequency \cite{key}. 
In the current set-up, the cavity coupling strength of the circular (rectangular) aperture at 110GHz is on the order of $10^{-2}$ ($10^{-4}$), which is considered to be in the weak coupling regime.
In this limit we believe that the RCM should continue to work, and it is valid to extend it in the manner described below. 
It will be shown later that the $Z_{in}$ statistics of a high-loss, weakly coupled cavity cascade system is dominated by the response from the first cavity \cite{Gradoni2012}. 
Despite this, we have recently utilized a machine learning algorithm to distinguish the number of cavities in the chain from raw $Z_{in}$ data \cite{Ma2019}.

In contrast to $Z_{in}$, the total cavity number plays a major role in the statistics of the trans-impedance $Z_{t}$ as shown in Fig. \ref{fig:Zinandt} (c) and (d) ($Re(Z_t)$ statistics in Fig. \ref{fig:supfull} (c)). The distribution of $Z_{t}$ displays progressively smaller values when the total number of connected cavities is increased. Note that with stronger coupling (circular aperture tests shown in Fig. \ref{fig:Zinandt} (c) vs. the rectangular aperture in Fig. \ref{fig:Zinandt} (d)), the measured $Z_{t}$ shows a larger chance to experience high impedance values compared with the weak coupling cases (the rectangular aperture has only 5 propagating modes).

\section{III. Model of coupled cavities}

Over the last decade, the Random Coupling Model has been introduced, studied extensively and compared with single chaotic enclosure experiments with various cavity losses, dimensions, and non-linear elements inside the cavity \cite{Gradoni2014,Xiao2018,Zhou2019}. Here, we apply an RCM-based model which can be used to make statistical predictions of impedance values in inter-connected systems of chaotic cavities. 
Our approach builds on an earlier work for single-mode connection between cavities \cite{Gradoni2012}. As shown schematically in Fig. \ref{fig:VNA}, our approach first adopts RCM to describe each individual chaotic enclosure \cite{Gradoni2014, Xiao2018}. The system-specific details are identified in yellow in Fig. \ref{fig:VNA}. These consist of the single-mode waveguide ports and the multi-mode apertures between cavities, and are described by the radiation impedance $Z_{port}$ and the radiation admittance matrix $\underline{\underline{Y_{aper}}}$, respectively. With known geometry, these frequency dependent complex coupling quantities can be determined through either full-wave numerical simulations or direct radiation measurements (see Appendix B. (b)). For an N-mode aperture we utilize the aperture admittance matrix $\underline{\underline{Y_{aper}}}$ (an $N\times N$ matrix) to describe its deterministic properties as a function of frequency. We use $Z_{port}$ to represent the deterministic properties of the single-mode ports in a manner identical to previous treatments of the port radiation impedance \cite{Zhou2017,Xiao2018}. 

An important consideration is the number of aperture modes (both propagating and evanescent) to include in $\underline{\underline{Y_{aper}}}$. The convergence of the impedance statistical prediction with the total number of included aperture modes is investigated in Fig. \ref{fig:conver}. The single cavity radiation admittance matrix can be written as a block-diagonal complex and frequency-dependent matrix
\begin{equation} \label{eq:yrad}
    \underline{\underline{Y_{rad}}}=\left[ \begin{matrix} \underline{\underline{Y_{rad,a}}} & 0 \\ 0 & \underline{\underline{Y_{rad,b}}}\end{matrix} \right]
\end{equation}
The $\underline{\underline{Y_{rad}}}$ subscripts ``$a$'' and ``$b$'' refer to the connection on the left and right sides of the cavity in the linear chain, while the values solely depend on the specific cavity connection (i.e., a port or aperture that allows $N_a$ modes in ``$a$'' and $N_b$ modes in ``$b$''). The off-diagonal zeros reflect the assumption that the apertures and ports are sufficiently separated such that no direct connection exists between them. This assumption can be lifted if direct illumination exists between apertures and ports \cite{Gradoni2015}. The RCM single cavity admittance matrix $\underline{\underline{Y_{cav}}}$ is then constructed as shown in Eqn. \ref{eq:ycavv} from $\underline{\underline{Y_{rad}}}$ and the $(N_a+N_b)\times(N_a+N_b)$ dimensionless universal fluctuation matrix $\underline{\underline{\xi_{RCM}}}$ whose statistics are determined solely by the loss parameter $\alpha$ of the cavity,
\begin{equation} \label{eq:ycavv}
    \underline{\underline{Y_{cav}}}=i\cdot Im\left(\underline{\underline{Y_{rad}}}\right) \, + \, Re\left(\underline{\underline{Y_{rad}}}\right)^{0.5}\cdot \underline{\underline{\xi_{RCM}}}\cdot Re\left(\underline{\underline{Y_{rad}}}\right)^{0.5}
\end{equation}
The matrix $\underline{\underline{\xi_{RCM}}}$ can be calculated using a Monte-Carlo technique \cite{Zheng2006,Zheng2006a}. The single cavity admittance matrix $\underline{\underline{Y_{cav}}}$ reflects the chaotic universal fluctuations from $\underline{\underline{\xi_{RCM}}}$ \cite{Gradoni2014}, ``dressed'' by the system-specific properties of the ports and apertures described by $\underline{\underline{Y_{rad}}}$. 

For a description of the statistical properties of the cavity cascade, cavities can be connected together by enforcing continuity of voltages and currents at the cavity coupling planes. With this recipe, the complete RCM cavity cascade model is created. One can then make predictions for the statistics of $Z_{in}$ and $Z_t$ of the entire system (see Eqns. \ref{eq:zin}, \ref{eq:zt} in the Appendix A) based on the minimal information of cavity loss and system coupling.

Two conclusions are drawn from inspection of the $Z_t$ and $Z_{in}$ formulas: (1) with high cavity loss ($\alpha>1$) and weak inter-cavity coupling, the expression for $Z_{in}$ can be approximated as $Z_{in}\approx Z_{port}/\xi_{aa}^{(1)}$, where $Z_{port}$ is the radiation impedance of the waveguide port, and $\xi_{aa}^{(1)}$ is the diagonal component of the $\underline{\underline{\xi_{RCM}}}$ matrix of the first cavity. In this case, the statistics of $Z_{in}$ has a mean equal to $Z_{port}$ and a fluctuation determined solely by the loss parameter of the first cavity, consistent with the data in Fig. \ref{fig:Zinandt} (a) and (b), 
(2) as shown in Eqn. \ref{eq:zt}, the trans-impedance $Z_{t}$ of the multi-cavity system is expressed as the product of the system input impedance $Z_{in}$ and $N$ multipliers $\underline{\underline{\Gamma}}^{(i)}$, where $N$ is the total number of cavities. The elements of $\underline{\underline{\Gamma}}^{(i)}$ have small values in the high-loss cases (see Appendix A (c)).
Since the statistics of $Z_{in}$ remain essentially unchanged as discussed in (1), the addition of connected cavities will introduce extra $\underline{\underline{\Gamma}}^{(i)}$ as multipliers (see Appendix A (c)), and further result in $Z_t$ values closer to zero and having smaller fluctuations, as observed in the data shown in Fig. \ref{fig:Zinandt} (c) and (d).

\section {III. Comparison and discussion}

\begin{figure*}
\includegraphics[width=0.95\textwidth]{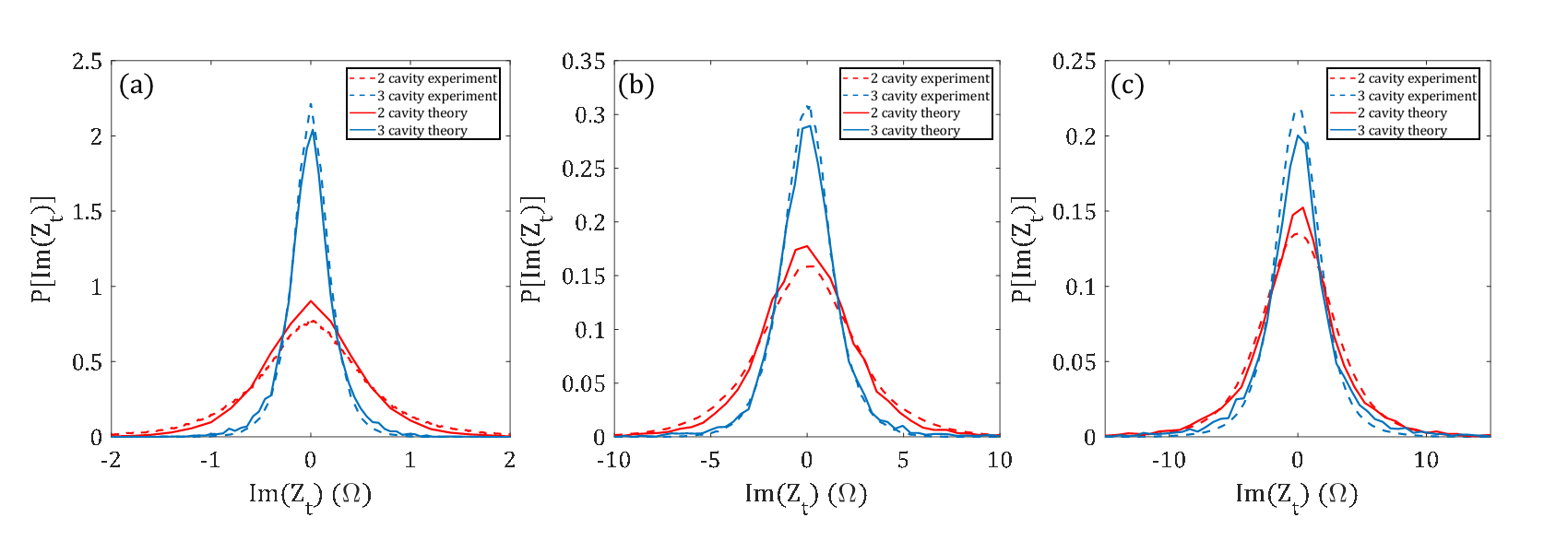}
\caption{\label{fig:compare1} Comparison of the imaginary part of trans-impedance ($Im(Z_t)$) between the experimental curves and the theoretical predictions for 2 and 3 cavity cascade system. The red curves are for two-cavities and blue are for three-cavities. Figure (a) and (b) are from scaled measurements with rectangular and circular shaped apertures, respectively. Figure (c) is from the full-scale system. The full scale analysis with different cavity loss is shown in Fig. \ref{fig:supfull}}
\end{figure*}

With the single cavity RCM loss parameters and system coupling detail presented, we are in position to compare impedance statistics from model predictions with those measured in experiments. 
The loss parameters, obtained from single cavity transmission measurements, are $\alpha=9.1$ and $\alpha=9.7$ for the scaled and full scale cavity with 6 absorber cones \cite{Xiao2018}, respectively. Since all the cavities inside the linear array are electrically identical (nominally), the same cavity loss parameter $\alpha$ will be assigned to each cavity in the cascade. The aperture and the port radiation admittances are obtained by numerical simulations in CST and direct measurements, respectively (see Appendix B. (b)). 
Combining this information, we generate an ensemble of 2 and 3-cavity cascade system impedances, and compare its statistics with those of the measured data. In Fig. \ref{fig:compare1}, we compare the statistics of $Im(Z_t)$ for cavity cascade systems with various cavity numbers, coupling strengths and physical dimensions.
Fig. \ref{fig:compare1} (a) show PDF's of $Im(Z_t)$ for 2 and 3 cavity systems when the cavities are connected by rectangular shaped apertures having 5 modes ($10^{-4}$ coupling strength). The statistics of the 2 and 3 cavity cascade theory generated $Im(Z_t)$ (lines) match well with the measured data (dashed lines). A minor mismatch between the 2-cavity $Im(Z_t)$ statistical comparison is observed, where the theory generated $Im(Z_t)$ has smaller values and is closer to zero as compared with the measured data.
The inter-cavity coupling strength is increased to $10^{-2}$ by employing the circular shaped aperture, and as shown in Fig. \ref{fig:compare1} (b), this increases the fluctuations of the trans-impedance for both the 2 and 3 cavity systems. Results for the full-scale measurements with circular aperture connections are shown in Fig. \ref{fig:compare1} (c). 
Good agreement between model and measurements are obtained in all cases. Similar to the rectangular aperture connection case, the theory tends to predict smaller(larger) $Im(Z_t)$ values for 2(3) cavity cascade systems with circular aperture connections.

Aside from validating the prediction performance of the RCM cavity cascade model, another key aspect of our experiment is to study the miniature cavity technique for the multi-cavity systems. As introduced in section I, the full scale cavity is built with linear dimensions 20 times larger compared with its scaled counterpart. With the operating frequency properly scaled and loss parameter made equal by adjusting the wall conductivity, the statistical wave properties of the two set-ups are expected to be identical \cite{Xiao2018}. The direct comparison of system trans-impedance statistics can be examined by comparing the $Im(Z_t)$ PDFs shown in Fig. \ref{fig:compare1} (b) and (c). The scaled and full scale 2 and 3 cavity $Im(Z_t)$ PDFs have similar peak values and spread. They differ in the exact PDF peak values and the relative differences between the 2 and 3 cavity $Im(Z_t)$ curves.
We believe that this imperfect agreement between scaled and full scale $Im(Z_t)$ PDFs is caused by a difference in the aperture thickness. The circular aperture thickness is about $\lambda_{op}$ in the scaled cavities, but only $0.04\lambda_{op}$ in the full scale set-up ($\lambda_{op}$ represents the characteristic operating wavelength). Note that the finite thickness of the apertures are included in the full-wave $\underline{\underline{Y_{rad}}}$ simulations, resulting in good agreement between model and measurements. By scaling the thickness of the full scale aperture to one $\lambda_{op}$, identical frequency dependent $\underline{\underline{Y_{rad}}}$ is observed as compared to the scaled aperture's. 
One can further conduct the multi-cavity RCM model calculations using such `scaled' full scale aperture $\underline{\underline{Y_{rad}}}$. The model generated full scale impedance statistics matched well with the scaled cases (Fig. \ref{fig:compare1} (b)).

\begin{figure}
\includegraphics[width=0.5\textwidth]{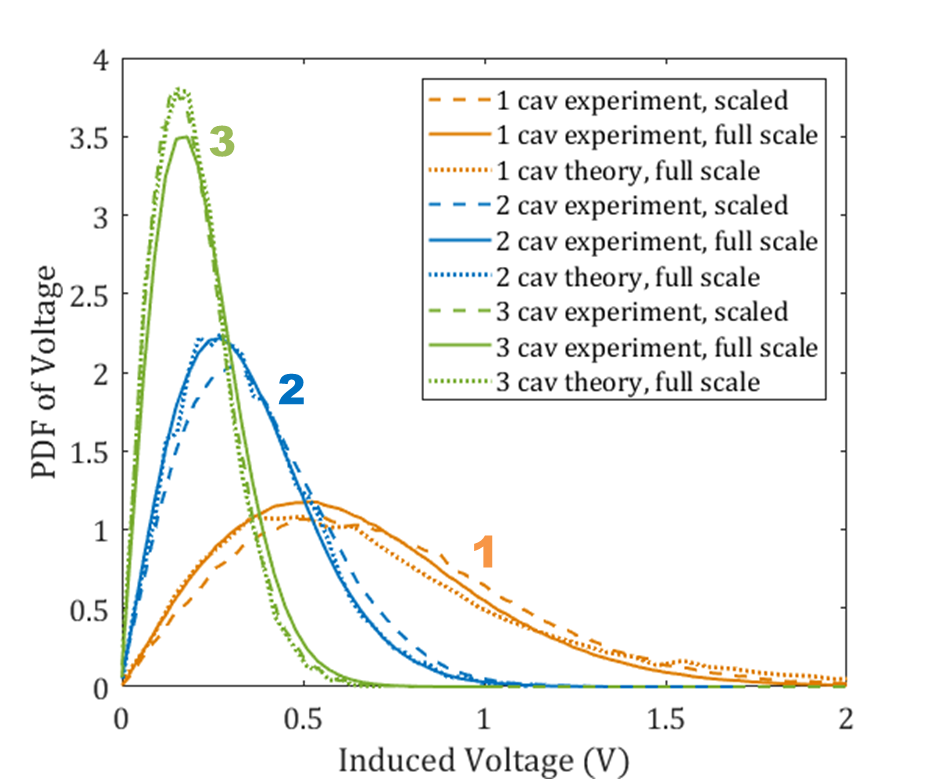}
\caption{\label{fig:inducedV} The PDFs of induced voltage on a 50 Ohm load attached to the last cavity of the full scale cavity cascade systems (3.95-5.85GHz) and its scaled counterparts (75-110GHz). The curves corresponding to the 1, 2 and 3 cavity system are color-coded in blue, yellow and green, respectively. The full scale system experimental (theoretical) data are shown at solid (dotted) lines, and the scaled experiment data are shown as dashed lines.}
\end{figure}

We are also able to calculate the statistics of the magnitude of the induced voltage delivered to a 50$\Omega$ load impedance attached to the last cavity in the 1D chain. The load induced voltage is calculated from the measured impedance based on the analysis presented in Refs. \cite{Hemmady2012,GilGil2016}. The model-generated induced voltage is calculated using Eqns. \ref{eq:zin} and \ref{eq:zt} in the appendix. The input powers used in the experimental and model generated induced load voltage are set as $1W$. 
Despite the differences in aperture thickness, good agreement between these two set-ups as found for the induced voltage statistics shown in Fig. \ref{fig:inducedV}. The results show that such a scaling technique can be very conveniently extended from single to multi-cavity systems, allowing investigation of systems with a large number of cavities and more sophisticated connection topology.

The proposed theoretical formulation is not expected to work at the extreme high-loss limits ($\alpha \rightarrow \infty$) due to the failure of the random plane wave hypothesis, which is a prerequisite of the RCM. 
This breakdown can be expected when the estimated 3dB Q-width of a mode becomes  comparable to operating frequency ($Q\sim 1$ or $\alpha \sim \omega/\Delta \omega \gg 1$).
However, the model is valid for moderately large loss ($\omega/\Delta \omega \gg \alpha \gg 1$), and the impedance statistics simplify to Gaussian distributions in the $\alpha \gg 1$ limit \cite{Zheng2006,Zheng2006a}. The RCM theoretical formulation can be applied to lower loss systems ($\alpha \sim 1$), but the stronger impedance fluctuations of low-loss systems poses great challenges for the acquisition of good statistical ensemble data by either numerical or experimental methods \cite{Yeh2013,Yeh2013a}. 
The formulation will also require modifications of the cavity total admittance matrix when the inter-cavity coupling is increased substantially. Non-zero off-diagonal components of the $\underline{\underline{Y_{rad}}}$ matrix (Eqn. \ref{eq:yrad}) must be determined when direct coupling is prominent. Failure to include these off-diagonal terms may contribute to the lack of detailed agreement between the model-generated and experiment results in Fig. \ref{fig:compare1}.

\section{V. Conclusion}

In this paper we report experimental results on the scattering properties of coupled multi-cavity complex systems and present a model based on the Random Coupling Model to generate an ensemble of impedance predictions whose statistical properties agree well with the measurements. The proposed model holds for all scenarios considered in this paper, including varying cavity number, cavity loss and inter-cavity coupling. The frequency and dimensional scaling technique is also expanded from the single cavity case \cite{Xiao2018} to the multi-cavity regime, opening up new possibilities to study electrically large coupled cavity systems in a convenient laboratory setting. The proposed experimental and theoretical methods should be useful for analyzing coupled physical systems whose components exhibit wave chaotic behavior. Examples include conductance fluctuations of coupled quantum dot systems where the single-electron dynamics in the dot are chaotic \cite{RevModPhys.69.731}, and EM properties such as the shielding effectiveness and power flow patterns of a coupled enclosure system. In future work, we will explore the crossover in system impedance statistics as the enclosures go from weakly coupled to strongly coupled, and investigate systems with more complex connection topology.

\begin{acknowledgments}
This work was supported by ONR under Grant No. N000141512134, ONR Grant No. N000141912481, ONR Grant No. N0001418WX00096, ONR Grant No. N0001418WX00038, ONR Grant No. N0001418WX00053, ONR Grant No. N0001419WX00518, AFOSR COE Grant FA9550-15-1-0171, and ONR DURIP Grant N000141410772.
\end{acknowledgments}

\appendix*
\section{Appendix A: RCM cavity cascade model}

\begin{figure}
\includegraphics[width=0.5\textwidth]{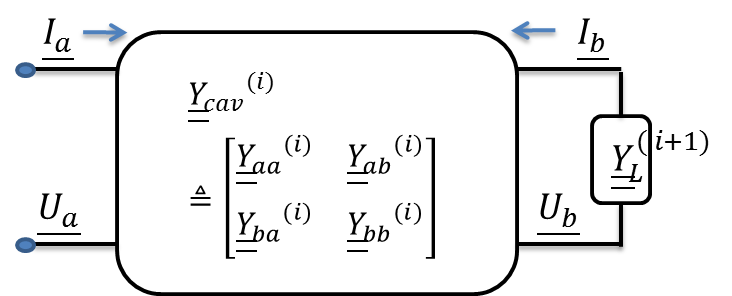}
\caption{\label{fig:casunit} Schematic diagram of the $i^{th}$ cavity in a cascade with part $a$ on the left and part $b$ on the right. The U and I refers to the voltage and current at two sides of the cavity.}
\end{figure}

Here we give a brief derivation of expressions for the input and trans-impedances of a coupled multi-enclosure system through extension of the RCM.

(a) Single Cavity Treatment

We write the radiation admittance matrix $\underline{\underline{Y_{rad}}}$ of a single cavity, as:

\begin{equation}
    \underline{\underline{Y_{rad}}}=\left[ \begin{matrix} \underline{\underline{Y_{rad,a}}} & 0 \\ 0 & \underline{\underline{Y_{rad,b}}}\end{matrix} \right]
\end{equation}

The $\underline{\underline{Y_{rad}}}$ is a block diagonal matrix, where the diagonal elements $\underline{\underline{Y_{rad, a}}}$ or $\underline{\underline{Y_{rad, b}}}$ are either the radiation admittance of the port $Y_{rad, port}$ or the radiation admittance matrix of the aperture modes $\underline{\underline{Y_{rad, aper}}}$, depending solely on whether a port or an aperture is connected to that side of the cavity. The off-diagonal components are set to zero under the assumption that there is no direct coupling between ports and apertures.
We next incorporate the RCM fluctuating quantities into the description of the fluctuating cavity admittance. The cavity admittance matrix $\underline{\underline{Y_{cav}}}$ is:

\begin{equation}
    \underline{\underline{Y_{cav}}}=i\cdot Im\left(\underline{\underline{Y_{rad}}}\right) \, + \, Re\left(\underline{\underline{Y_{rad}}}\right)^{0.5}\cdot \underline{\underline{\xi}}\cdot Re\left(\underline{\underline{Y_{rad}}}\right)^{0.5}
\end{equation}
We represent the cavity admittance matrix $\underline{\underline{Y_{cav}}}$ (as shown in Fig. \ref{fig:casunit}) as
\begin{equation}
    \underline{\underline{Y_{cav}}} = \left[ \begin{matrix}  \underline{\underline{Y_{aa}}} & \underline{\underline{Y_{ab}}} \\ \underline{\underline{Y_{ba}}} & \underline{\underline{Y_{bb}}}  \end{matrix}\right]
\end{equation}
We then have:
\begin{equation}
    \left[ \begin{matrix} \underline{I_a} \\ \underline{I_b} \end{matrix}\right]=\underline{\underline{Y_{cav}}} \, \left[ \begin{matrix} \underline{U_a} \\ \underline{U_b} \end{matrix}\right]=\left[ \begin{matrix}  \underline{\underline{Y_{aa}}} & \underline{\underline{Y_{ab}}} \\ \underline{\underline{Y_{ba}}} & \underline{\underline{Y_{bb}}}  \end{matrix}\right]\, \left[ \begin{matrix} \underline{U_a} \\ \underline{U_b} \end{matrix}\right]
\end{equation}
where the matrix $\underline{\underline{\xi}}$ is the normalized impedance which has the universal
statistical properties predicted by Random Matrix Theory (depending only on the loss parameter $\alpha$). The dimension of the matrix $\underline{\underline{\xi}}$ is equal to the dimension of $\underline{\underline{Y_{rad}}}$, which is $N_a+N_b$, where $N_a$ or $N_b$ is the dimension of the radiation admittance matrices on the two sides of this cavity.

(b) Total Admittance of an N-Cavity Chain

After we construct the $\underline{\underline{Y_{cav}}}$ matrices for all cavities in the cascade, we are in the position of developing the cascade quantities for the enclosure chain. For the $i^{th}$ cavity in the cavity cascade chain, with the information of how the $i^{th}$ cavity is connected to its left (previous) and right (next) neighboring cavities, and knowledge of the load admittance presented by the $(i+1)^{th}$ cavity $\underline{\underline{Y_L^{(i+1)}}}$ and everything beyond it, we will have an iterative approach to calculating the total chain admittance matrix:

\begin{equation} \label{eq:cont}
    \left\{  \begin{matrix}
         \left[ \begin{matrix} \underline{I_a} \\ \underline{I_b} \end{matrix}\right]=\underline{\underline{Y}}^{(i)}\cdot\left[ \begin{matrix} \underline{U_a} \\ \underline{U_b} \end{matrix}\right] \\
         -\underline{I_b}=\underline{\underline{Y_L}}^{(i+1)}\cdot \underline{U_b}
    \end{matrix} \right.
\end{equation}

Eqn. \ref{eq:cont} expresses the continuity of voltage and current in all modes of the aperture. By solving Eqn. \ref{eq:cont} in matrix form, we have: 

\begin{equation} \label{eq:ubua}
    \underline{U_b}=-\left(\underline{\underline{Y_{bb}}}^{(i)}+\underline{\underline{Y_L}}^{(i+1)}\right)^{-1}\,\underline{\underline{Y_{ba}}}^{(i)}\,\underline{U_a}
\end{equation}
\begin{equation} \label{eq:yl}
    \underline{\underline{Y_{L}}}^{(i)}=\underline{\underline{Y_{aa}}}^{(i)}-\underline{\underline{Y_{ab}}}^{(i)}\,\left(\underline{\underline{Y_{bb}}}^{(i)}\,+\,\underline{\underline{Y_{L}}}^{(i+1)}\right)^{-1}\underline{\underline{Y_{ba}}}^{(i)}
\end{equation}

Eqn. \ref{eq:ubua} connects the voltages on two sides of a cascade unit, and Eqn. \ref{eq:yl} gives the $\underline{\underline{Y_L}}$ recursion relationship which calculates the load admittance $\underline{\underline{Y_L}}^{(i)}$ of the $i^{th}$ cavity given the $\underline{\underline{Y_L}}^{(i+1)}$ of the $(i+1)^{th}$ cavity and everything beyond it. At the end of the cavity chain, the load impedance is the VNA load impedance $Z_0=50\Omega$. We can calculate the total load admittance of the entire structure from the end load `forward' to the input port at the first cavity.

(c) Input and Trans-Impedances

We will investigate the trans-impedance $Z_t$ and the input impedance $Z_{in}$ of the cascaded system from the voltage and $Y_L$ recursion relationships. For the scalar $Z_{in}$, we have

\begin{equation} \label{eq:zin}
    Z_{in}=\frac{U_a^{(1)}}{I_a^{(1)}}=\frac{1}{Y_L^{(1)}}
\end{equation}
where the $U_a^{(1)}$ and $I_a^{(1)}$ refers to the scalar voltage and current at the input side of the first cavity. 

From Eqn. \ref{eq:ubua} and \ref{eq:yl}, we have:

\begin{equation} \label{eq:zt}
\begin{aligned}
    Z_t=\frac{U_b^{(N)}}{I_a^{(1)}} &=\frac{-\left({Y_{bb}}^{(N)}+{Y_L}^{(N+1)}\right)^{-1}\,\underline{{Y_{ba}}}^{(N)}\,\cdot \,\underline{U_a^{(N)}}}{I_a^{(1)}}\\
    &=\frac{-\left({Y_{bb}}^{(N)}+{Y_L}^{(N+1)}\right)^{-1}\,\underline{{Y_{ba}}}^{(N)}\,\cdot \,\underline{U_b^{(N-1)}}}{I_a^{(1)}} \\
    &=\prod_{i=1}^N\left[-\left(\underline{\underline{{Y_{bb}}}}^{(i)}+\underline{\underline{{Y_{L}}}}^{(i+1)}\right)^{-1}\,\underline{\underline{{Y_{ba}}}}^{(i)} \right] \cdot \frac{U_a^{(1)}}{I_a^{(1)}}\\
    &=\prod_{i=1}^N\left[\underline{\underline{\Gamma}}^{(i)}\right] \cdot Z_{in}
    \end{aligned}
\end{equation}

The $U_b^{(N)}$ is the scalar voltage at the load connected to the output side of the last ($N^{th}$) cavity.
The multiplier $\underline{\underline{\Gamma}}^{(i)}$ of the $i^{th}$ cavity is defined as $\underline{\underline{\Gamma}}^{(i)}=-\left(\underline{\underline{{Y_{bb}}}}^{(i)}+\underline{\underline{{Y_{L}}}}^{(i+1)}\right)^{-1}\,\underline{\underline{{Y_{ba}}}}^{(i)}$. 
$\underline{\underline{\Gamma}}^{(i)}$ can be varied from a matrix or a vector depending on the specific location of the $i^{th}$ cavity in the entire cascade chain (e.g., $\underline{\underline{\Gamma}}^{(i)}$ is a matrix for $i$ in $2\sim N-1$, and a vector ($\underline{\Gamma}^{(i)}$) at the first and the last cavity).
With Eqn. \ref{eq:zin} and \ref{eq:zt}, we present the full theoretical formulas of $Z_{in}$ and $Z_t$ based on the RCM. 

\section{Appendix B: Experimental Details, Aperture Admittance Calculations, and Additional Data}

(a) Cavity cascade experimental set-up

\begin{figure}
\includegraphics[width=0.5\textwidth]{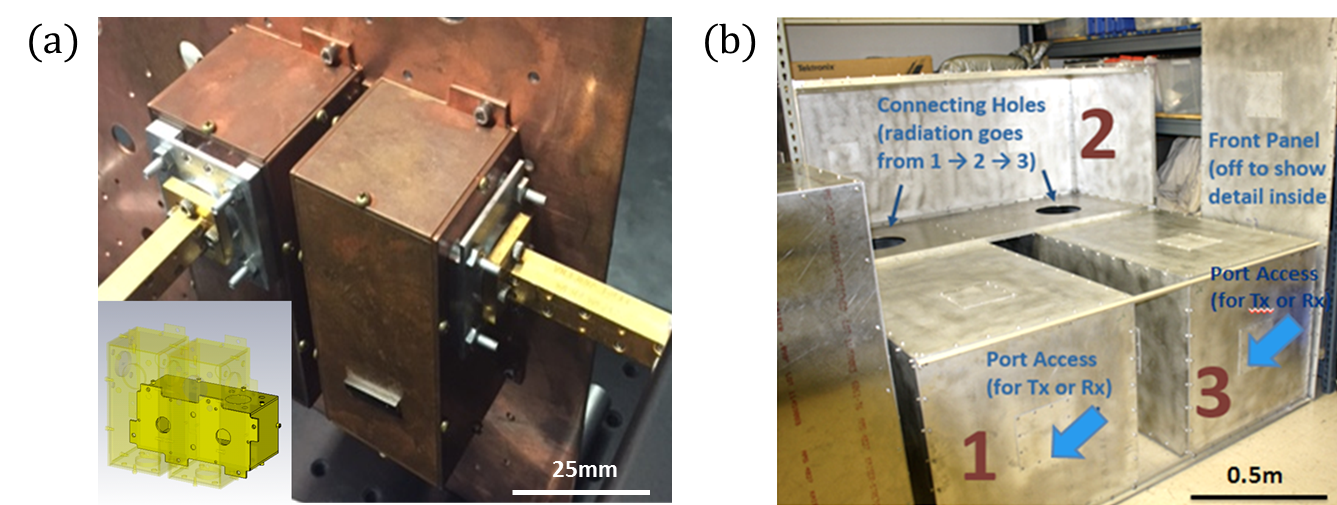}
\caption{\label{fig:connec} (a) The scaled 3 cavity cascade case (where the second one is hidden behind the vertical copper plate) connected to single-mode WR10 waveguides. The inset shows how the second cavity is arranged and the location of the two circular apertures. (b) Picture of the full scale cavity cascade set-up with one wall of cavity 2 removed. The cavities are connected through circular apertures. Individual perturbers and RF absorber cones are employed inside the enclosures (not shown).}
\end{figure}

The details of the cascade cavity experimental set-up are shown in Fig. \ref{fig:connec}. The scaled and full scale 3-cavity cascade structures are shown in Fig. \ref{fig:connec} (a) and (b), respectively. The scaled experiments are conducted in University of Maryland, and the full scale version at Naval Research Laboratory. With the single cavity losses made equal and the physical dimensions carefully scaled, nominally identical electromagnetic
conditions are achieved between the two set-ups.

(b) Aperture and port radiation admittance studies

The radiation admittance $Y_{rad}$ and impedance $Z_{rad}$ refer to the cases where apertures or ports radiate into free space \cite{Yeh,Gradoni2015}. Numerical simulations and experimental methods are adopted in order to characterize the radiation information of the apertures and ports employed in the experimental set-ups. 

\begin{figure}
\includegraphics[width=0.5\textwidth]{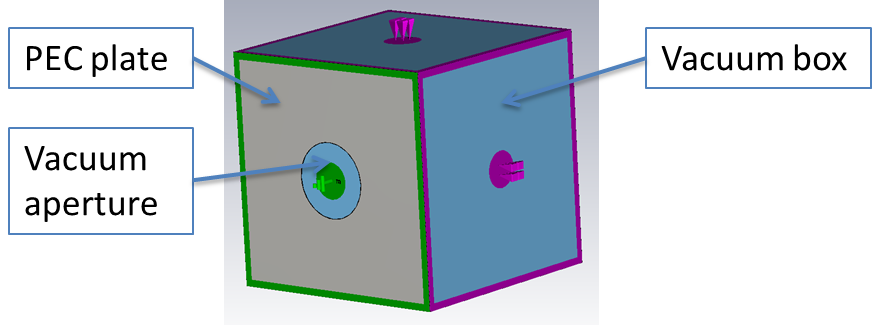}
\caption{\label{fig:CST1} The model set-up for aperture admittance matrix calculations in the CST simulations. A PEC plate is carved with the dimension of the aperture, and attached to a vacuum box. The port is assigned at the 2D surface of the aperture and the total number of port modes is input into the CST solver.}
\end{figure}

We use Computer Simulation Technology (CST) \cite{CST} to calculate the frequency dependent radiation admittance matrix of both the scaled and full scale apertures. As shown in Fig. \ref{fig:CST1}, the aperture is carved out of the surface of a Perfect Electrical Conductor (PEC) plate. The thickness of the PEC plate equals to the thickness of the aperture of interest. The carved plate is attached to a large vacuum box whose boundaries are assigned as radiating. The simulation is run in the eigenvalue solver mode. Given the operating frequency and the total number of aperture modes, the simulation calculates the frequency dependent complex admittance matrix of the aperture. Similar techniques can be found in this aperture cross section numerical study of Ref. \cite{Gunnarsson2014}. 

The radiation impedance of the ports are obtained from experimental methods. For the WR10 waveguide port used in the scaled set-up, the $1\times 1$ radiation S-parameter ($S_{rad}$) is measured by putting the bare port inside a large radiating environment. The $Z_{rad}$ is then transformed from the measured $S_{rad}$. For the WR178 waveguide ports used in the full scale measurements, time-gating techniques are adopted to calculate the corresponding $Z_{rad}$ \cite{Addissie}.

(c) Convergence of the RCM cavity cascade formalism with aperture mode number

\begin{figure}
\includegraphics[width=0.4\textwidth]{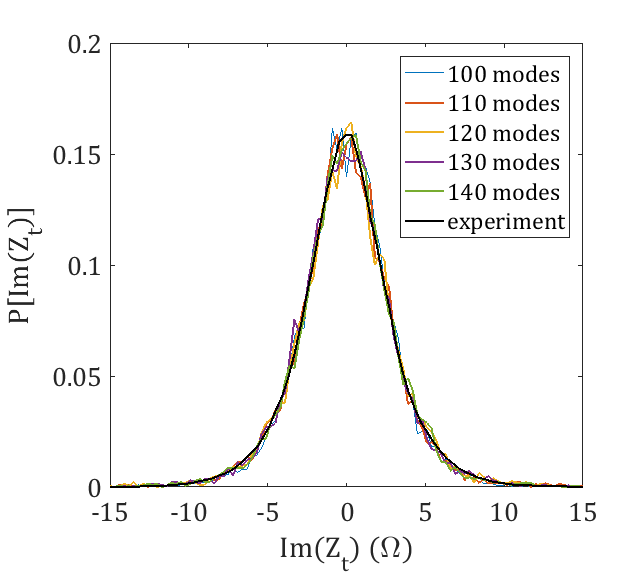}
\caption{\label{fig:conver} Statistics of the imaginary part of the trans-impedance for a scaled 3-cavity cascade with circular aperture connections, illustrating the convergence of the theoretical model with increasing mode number included in the aperture admittance matrix. The circular aperture allows 100 propagating modes. The PDFs of theory curves are unchanged when the non-propagating modes are included.}
\end{figure}

As introduced in the main text, the aperture radiation admittance $Y_{rad}$ is a $(n,n)$ matrix where $n$ is the total number of considered aperture modes. Though the total number of propagating modes can be calculated for given aperture dimensions and operating frequency, the choice of the total number of non-propagating modes which are taken into account is unclear. Here we validate that the RCM cavity cascade formalism can find convergence with an increasing number of considered non-propagating modes. As shown in Fig. \ref{fig:conver}, the convergence of the model is tested by adding circular aperture modes from 100 to 140 in the 3-cavity connection case. The statistics obtained from the theoretical model with different aperture mode numbers remains unchanged beyond 100 modes, and agrees well with the experimental results in that limit.

(d) Extended cavity cascade studies

\begin{figure*}
\includegraphics[width=0.9\textwidth]{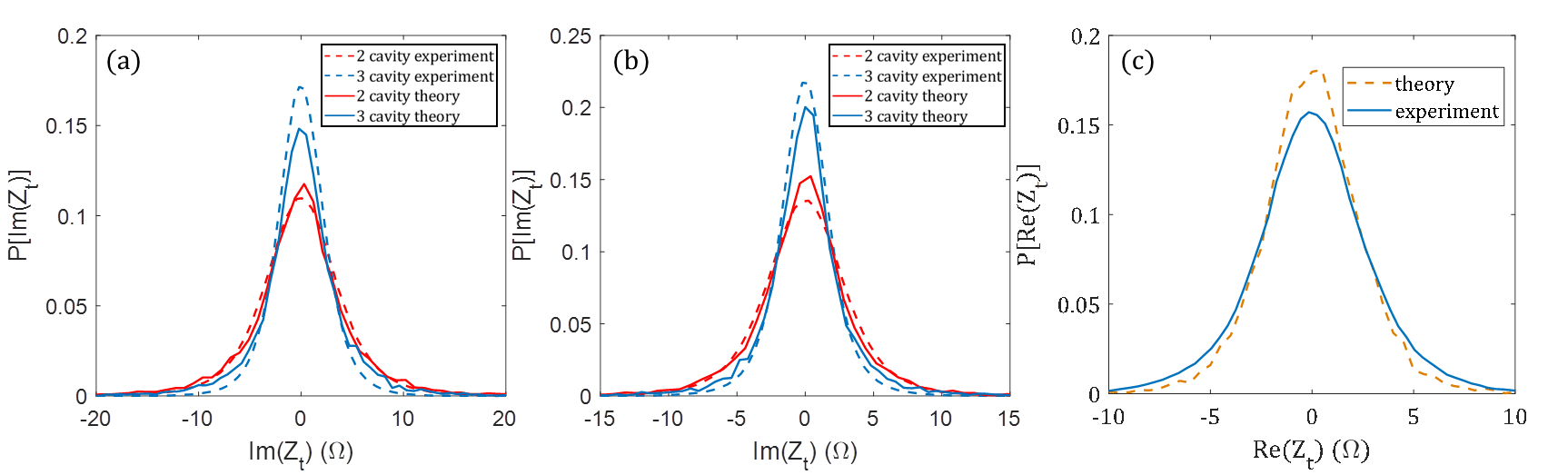}
\caption{\label{fig:supfull} (a) and (b): Statistics of the imaginary part of the trans-impedance of 2- and 3-cavity cascades when the loss of the system changes. (c): Comparison of the $Re(Z_t)$ statistics experimental and model results for the scaled 2 cavity set-up, with circular aperture connection. In (a), (b) and (c), the single cavity loss parameter $\alpha$ is measured as $\alpha=5.7$, $7.5$ and $9.1$, respectively.}
\end{figure*}

The 2- and 3-cavity cascade systems are studied with various single cavity loss parameter values. As shown in Fig. \ref{fig:supfull} (a) and (b), the statistics of 2- and 3-cavity model and experiment $Im(Z_t)$ are shown with single cavity loss parameter $\alpha=5.7$ and $7.5$, respectively. The cascaded cavity structure studied in Fig. \ref{fig:supfull} (a) and (b) are full scale structures with circular shaped aperture connections. By placing 2 and 4 RF absorber cones in each individual cavities, single cavity loss parameter are measured as $\alpha=5.7$ and $7.5$, respectively. We observed reasonably good agreement between the model-generated $Im(Z_t)$ PDFs (lines) and the measured results (dashed lines). In Fig. \ref{fig:supfull} (c), the statistics of the $Re(Z_t)$ of the scaled 2 cavity cascade with circular aperture connection is studied. We observed good agreement between the measured data and model-generated results.


\bibliographystyle{apsrev4-1} 

%

\end{document}